\documentclass[cameraready]{Interspeech}

\title{EChO-Agent: Evidence Chain Orchestration Agent for Audio Reasoning}

\author[affiliation={1}, equalcontribution, correspondingauthor]{Siyuan}{Zhang}
\author[affiliation={1}, equalcontribution]{Jian}{Zong} 
\author[affiliation={2}]{Junyu}{Wang}
\author[affiliation={3}]{Peiyuan}{Jiang}
\author[affiliation={4}]{Jiahao}{Yan}
\author[affiliation={5}]{Jingyu}{Zhang}
\author[affiliation={6}]{Tianrui}{Wang}
\author[affiliation={7}]{Xiaobao}{Wang}
\author[affiliation={8}]{Longbiao}{Wang}
\author[affiliation={9}]{Jianwu}{Dang}

\address{
    $^1$ School of Artificial Intelligence, Tianjin University, Tianjin, China
}

\email{\{zhang\_siyuan, zongjian\_66\}@tju.edu.cn}


\usepackage{comment}
\usepackage{makecell}

\begin{document}

\maketitle

\begin{abstract}
    While LALMs show promise on audio question answering, they fail to focus on question-relevant segments of audio and provide a clear, checkable reasoning process  when dealing with complex audio reasoning. Reinforcement learning and tool-augmented prompting can help models better relate questions to audio but lack a reliable way to understand, integrate, and self-verify audio segments. To address this gap, we present EChO-Agent, a modular agent framework that reformulates complex audio QA as a planning, tool execution, evidence integration, and answer verification workflow. Experiments on MMAR benchmark show EChO-Agent improves both accuracy and rubric scores over baseline and ablation studies show evidence integration is the key factor.
\end{abstract}

\section{Introduction}
    Audio reasoning requires models to interpret rich aural information (e.g., speaker intent, affective state, duration and frequency of event) and derive logical conclusions from perceptual signals \cite{lake2017building,wei2022chain}. This requires not only recognizing acoustic events but also extracting critical evidence and binding it to stepwise inference~\cite{zhifei2025audio_reasoner,ma2025audiocot}.
    
\begin{figure*}[t]
  \centering
  \includegraphics[width=\textwidth]{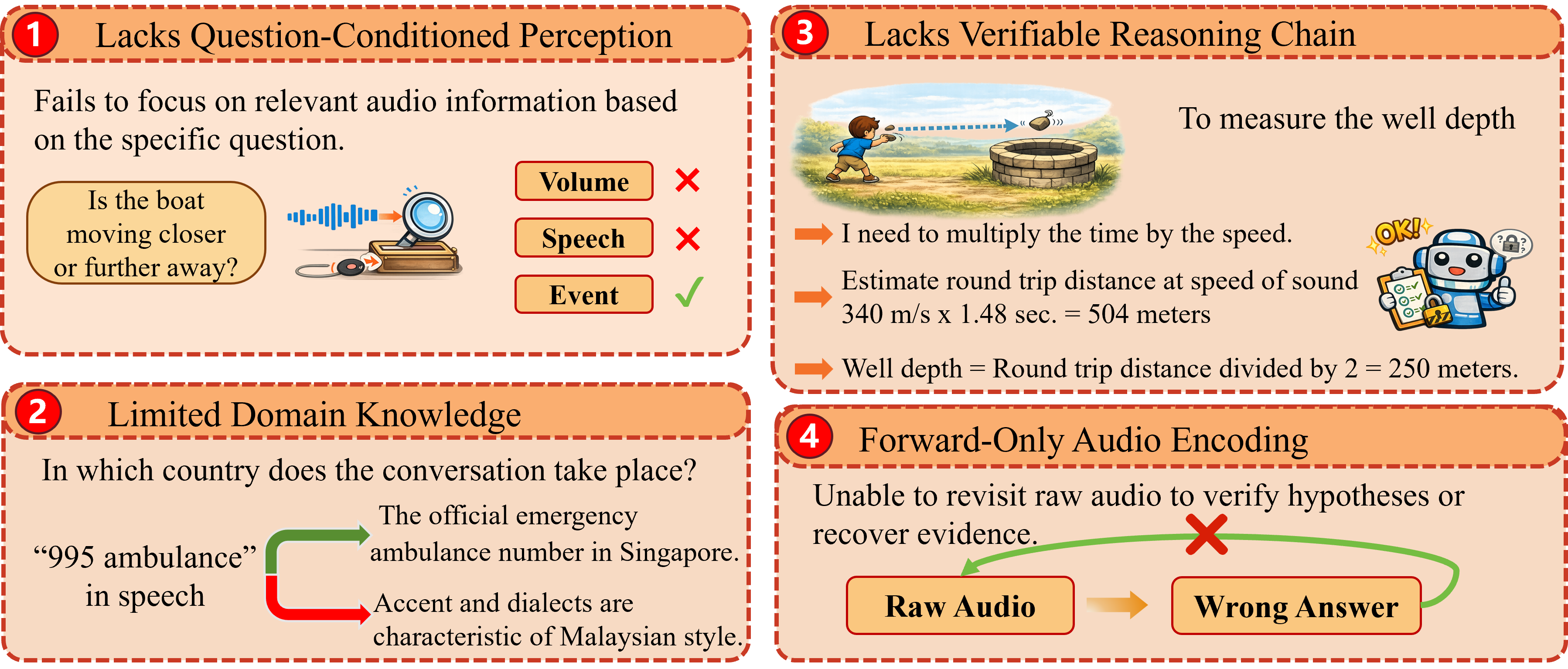}
  \caption{Key Limitations of Large Audio Language Models for Complex Audio Reasoning}
  \label{fig:Key Limitations}
\end{figure*}

    Large Audio Language Models (LALMs) integrate LLMs with audio encoders~\cite{tang2023salmonn,gong2023ltu,chu2023qwen}, showing strong potential for audio-text tasks ranging from captioning to question answering~\cite{xu2025qwen25omni,qwen2025qwen3omni,comanici2025gemini}, but complex audio reasoning remains unreliable due to several structural limitations~\cite{sakshi2024mmau,kong2024audioflamingo}. As summarized in Figure~\ref{fig:Key Limitations}, current LALMs lack question-conditioned perception, verifiable reasoning chains~\cite{wang2023selfconsistency,turpin2023language}, sufficient domain knowledge, and the ability to revisit audio to recover missed signals~\cite{cofiagent2026}. These issues are particularly consequential when models are evaluated not only by answer correctness but also by whether the reasoning process is faithful to audio evidence~\cite{lightman2023lets,sakshi2024mmau}. This concern is explicitly emphasized by the Interspeech 2026 Audio Reasoning Challenge, which evaluates \emph{process quality} using instance-level rubrics~\cite{ma2026interspeech_arc}. A seemingly good chain-of-thought of correct answer can still be penalized if it is weakly grounded or logically inconsistent with audio~\cite{turpin2023language,lyu2023faithful}.
    
    To address these challenges, we propose evidence chain orchestration agent for audio reasoning, short as EChO-Agent, a self-verifying tool agent that orchestrates a checkable evidence chain for audio reasoning. Our method decomposes inference into a four-stage pipeline: \textbf{Tool $\rightarrow$ Evidence $\rightarrow$ Reason $\rightarrow$ Verify}. First, agent arranges specialized audio tools to produce observations, then distills them into compact, logically structured evidence that tied to the question, after that agent performs reasoning grounded on raw audio and evidence, and finally verifies format compliance and evidence--answer consistency with dual-pass arbitration. Experiments on the MMAR benchmark demonstrate that EChO-Agent achieves \textbf{71.0\%} accuracy and a \textbf{63.0} rubric score, improving over the Qwen3-Omni baseline by \textbf{+2.3} accuracy points and \textbf{+4.3} rubric points, ranking \textbf{5th} on the MMAR Agent Track. Ablations confirm that structured evidence is the key factor: removing it causes the largest drop, even below the tool-free baseline, while the verification stage reduces avoidable last-mile failures.

\section{Related Work}
\label{sec:related}
\subsection{Audio Reasoning Challenge}

    The 2026 Interspeech Audio Reasoning Challenge~\cite{ma2026interspeech_arc} aims to evaluate \emph{process quality} rather than merely focusing on answer performance, requiring that the reasoning path be faithful to audio evidence. 
    Existing solutions, whether single-model or agent-based, often produce template-like reasoning, which appears coherent but has a weak foundation. The interpretability of intermediate evidence is limited, and the handling of tool failures is also ad hoc.
    These limitations suggest that complex audio reasoning should be treated as a principled evidence-seeking and checking process, not a pattern-matching exercise.

\subsection{Large Audio Language Models}

    Large Audio Language Models (LALMs), like LTU~\cite{gong2023ltu} and SALMONN~\cite{tang2023salmonn} align audio encoders with LLM backbone to support acoustic reasoning. Recent models like Qwen-Audio~\cite{chu2023qwen} inherit strong general-purpose priors and thinker-talker structure reduces explicit dependence on text modality, but audio tokens lack the discrete, interpretable structure needed for step-by-step inference. without a mechanism to identify which part of the audio matters for a specific query, models tend to rely on dominant acoustic patterns or linguistic priors, resulting in shortcut reasoning rather than grounded evidence~\cite{geirhos2020shortcut,turpin2023language}.Moreover, LALMs also operate in a closed loop after encoding: once subtle cues are missed, the model cannot revisit raw audio to verify hypotheses or recover evidence. 
    In summary, these limitations of current LALMs are exactly the reasons agentic frameworks were introduced in the first place: to augment foundation models with controllable reasoning and evidence-grounded execution.

\subsection{Tool-augmented Audio Agents}

    Recent works augment LALMs with external tools~\cite{schick2024toolformer} and optimize how tools are invoked. AuTAgent learns a tool-selection policy via reinforcement learning~\cite{autagent2026}; AudioRouter decouples tool routing from the reasoning model~\cite{audiorouter2026}; CoFi-Agent adds coarse-to-fine uncertainty-triggered re-analysis; and AudioRAG extends agentic pipelines with external knowledge retrieval for multi-hop reasoning~\cite{cofiagent2026,audiorag2026}. While these works improve acoustic perception, they primarily address what information is acquired, not how it is used. Raw tool outputs are typically passed directly to the reasoner without question-conditioned filtering, which can introduce distracting context and impede grounded inference~\cite{lyu2023faithful}. More importantly, none of these approaches provide a mechanism to synthesize tool outputs into concise, decision-critical evidence and verify the consistency between evidence and the final answer.

\begin{figure*}[t]
  \centering
  \includegraphics[width=\textwidth]{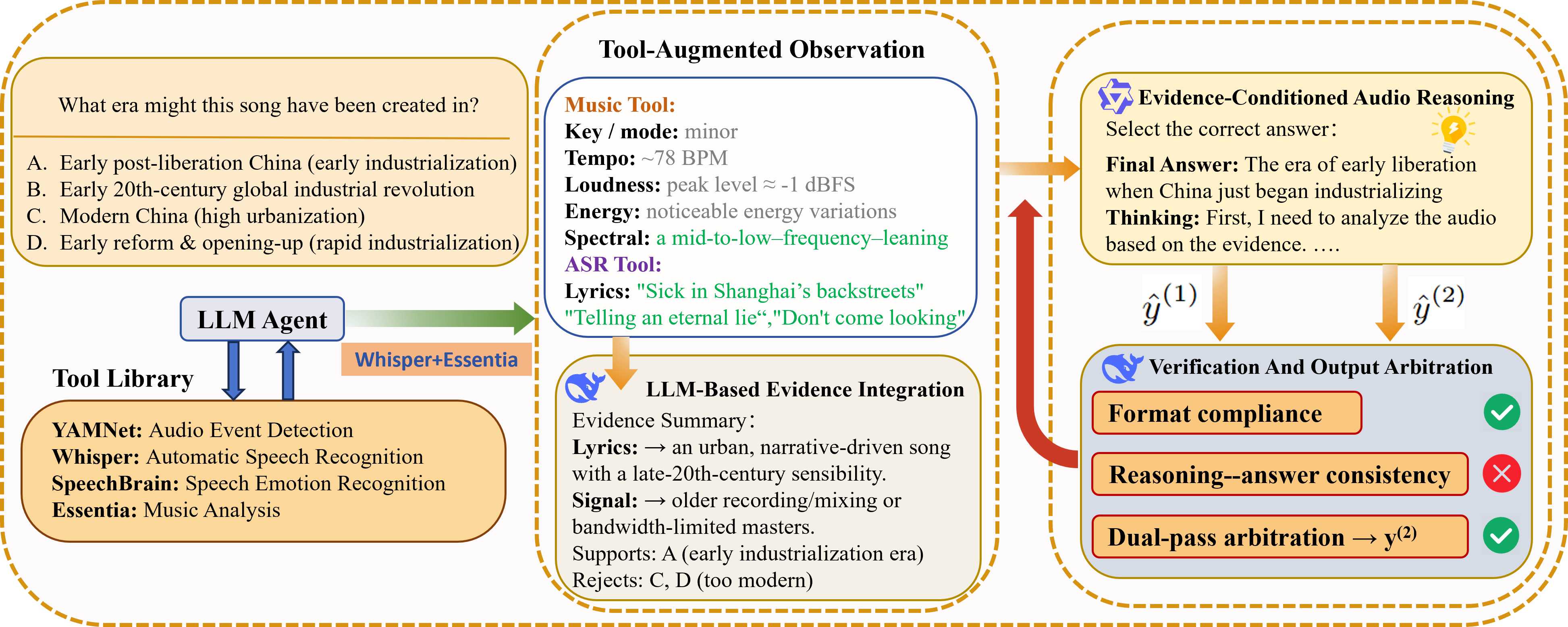}
  \caption{Auditable Tool-Augmented Agent Workflow for Audio Reasoning}
  \label{fig:workflow}
\end{figure*}

\section{Methodology}

\subsection{Task Formulation}

    Given an audio signal \(\mathbf{a}\) and a natural-language question \(q\) with candidate options \(\{c_1, \ldots, c_n\}\), the audio reasoning task requires producing the correct answer \(y^*\) along with a faithful chain-of-thought \(\hat{\mathcal{Y}}_{\text{CoT}}\) that is evaluated against instance-level rubrics. Directly prompting a Large Audio Language Model (LALM) \(\mathcal{M}_{\text{A}}\) to jointly perceive and reason often results in weakly grounded or shortcut reasoning, as discussed in Section~\ref{sec:related}. To address this, we decompose the task into a four-stage pipeline (Figure~\ref{fig:workflow}):
    \begin{align}
        (q, \mathbf{a}) \xrightarrow{\text{1.Tool}} \mathcal{O}
        \xrightarrow{\text{2.Evidence}} \mathcal{E}
        \xrightarrow{\text{3.Reason}} \hat{y}
        \xrightarrow{\text{4.Verify}} y^{*},\hat{\mathcal{Y}}_{\text{CoT}}
        \label{eq:pipeline}
    \end{align}
    Here, a LLM-based orchestrator agent \(\mathcal{M}_{\text{L}}\) takes the question \(q\) (with options \(\mathbf{a}\)), dispatches tool calls to obtain observations \(\mathcal{O}\), distills \(\mathcal{O}\) into evidence \(\mathcal{E}\), and manages verification; the LALM \(\mathcal{M}_{\text{A}}\) performs audio-grounded reasoning conditioned on the raw audio and \(\mathcal{E}\) to produce a candidate answer \(\hat{y}\), which is then verified into the final answer \(y^{*}\) with an auditable reasoning trace \(\hat{\mathcal{Y}}_{\text{CoT}}\).

\subsection{Tool-Augmented Observation}

    The first stage constructs an observation set \(\mathcal{O}\) by invoking a suite of audio analysis tools on \(\mathbf{a}\). We adopt a \emph{question-type-conditioned static dispatch} strategy: the orchestrator analyzes \(q\) and selects a predefined tool combination based on the question, eliminating variance from stochastic tool selection. The tool suite comprises four classes: \textbf{Audio Event Detection} (\(t_{\text{aed}}\)), where we use YAMNet~\cite{gemmeke2017audioset} to produce frame-level sound event labels with confidence scores, providing evidence about acoustic events and temporal occurrence; \textbf{Automatic Speech Recognition} (\(t_{\text{asr}}\)), where we adopt Whisper~\cite{radford2023robust} to transcribe spoken content when reasoning depends on linguistic cues; \textbf{Speech Emotion Recognition} (\(t_{\text{ser}}\)), where we use a SpeechBrain~\cite{ravanelli2021speechbrain} SER model (wav2vec2-based emotion classifier) to estimate the speaker's affective state with confidence; and \textbf{Music Analysis} (\(t_{\text{music}}\)), where we employ Essentia~\cite{bogdanov2013essentia} with common MIR extractors (e.g., tempo/key/beat) to derive measurable musical attributes supporting reasoning about music-related cues. For each tool \(t_j\), the orchestrator rephrases \(q\) into a tool-appropriate directive \(p_j\), yielding \(o_j = t_j(\mathbf{a}, p_j)\); failed invocations are retried (up to 2 times) and, if unrecoverable, recorded as \texttt{[UNAVAILABLE]} to prevent hallucination.

\subsection{LLM-Based Evidence Integration}
\label{sec:evidence}

Raw tool observations are typically verbose and may contain information irrelevant to \(q\). This stage employ \textbf{DeepSeek-V3}~\cite{deepseek2024v3} as the evidence constructor to distill \(\mathcal{O}\) into a compact evidence set \(\mathcal{E}\) that directly supports answering the question.

Unlike simply prompting an LLM to summarize or concatenate tool outputs, the evidence integration is guided by a structured instruction prompt that defines three operations:
~\textbf{Relevance filtering}: \(\mathcal{M}_{\text{L}}\) identifies question-relevant information and discards tangential content;
~\textbf{Cross-observation synthesis}: when multiple tools report overlapping information, \(\mathcal{M}_{\text{L}}\) consolidates them and resolves conflicts by comparing confidence levels or specificity;
~\textbf{Evidence structuring}: \(\mathcal{M}_{\text{L}}\) organizes the retained signals into a compact evidence chain that groups and orders the decision-critical facts in the way they will be used to answer the question. 

This stage serves as the \emph{bridge} between perception and reasoning: it converts raw multi-tool outputs into a structured, question-focused representation that the LALM can directly consume, reducing the burden on the LALM and enabling it to focus on evidence-grounded inference rather than noisy observation parsing.

\begin{table*}[htbp]
\centering
\caption{Omni Language Models (OLMs) Performance Comparison}
\setlength{\tabcolsep}{2.5pt}
\footnotesize
\resizebox{\textwidth}{!}{
\begin{tabular}{lccccccccc}
\toprule
\textbf{Models} & \textbf{Size} & \multicolumn{3}{c}{\textbf{Single Modality (\%)}} & \multicolumn{4}{c}{\textbf{Mixed Modalities (\%)}} & \textbf{Avg (\%)} \\
\cmidrule(r){3-5} \cmidrule(lr){6-9}
& & Sound & Music & Speech & \makecell{Sound-\\Music} & \makecell{Sound-\\Speech} & \makecell{Music-\\Speech} & \makecell{Sound-Music-\\Speech} & \\
\midrule
AnyGPT-chat~\cite{zhan2024anygpt} & 8B & 24.24 & 19.42 & 22.11 & 27.27 & 27.52 & 26.83 & 29.17 & 23.70 \\
OpenOmni~\cite{luo2025openomni} & 8B & 20.61 & 22.33 & 35.37 & 18.18 & 27.06 & 23.17 & 25.00 & 27.00 \\
Baichuan-Omni-1.5~\cite{li2025baichuan} & 11B & 41.21 & 33.01 & 40.48 & 36.36 & 48.62 & 39.02 & 41.67 & 40.70 \\
Qwen-2.5-Omni~\cite{xu2025qwen25omni} & 3B & 53.94 & 46.12 & 53.74 & 36.36 & 60.09 & 57.32 & 58.33 & 53.80 \\
Qwen-2.5-Omni & 7B & 58.79 & 40.78 & 59.86 & 54.55 & 61.93 & 67.07 & 58.33 & 56.70 \\
Gemini 2.0 Flash~\cite{comanici2025gemini} & -- & 61.21 & 50.97 & 72.11 & \textbf{\underline{81.82}} & 72.48 & 65.85 & 70.83 & 65.60 \\
Qwen-3-Omni-instruct~\cite{qwen2025qwen3omni} & 30B & 58.79 & 52.91 & 78.23 & 72.73 & \textbf{\underline{77.06}} & 70.73 & 70.83 & 68.70 \\
Qwen-3-Omni-thinking & 30B & \textbf{\underline{66.67}} & 49.51 & 79.93 & 63.64 & 74.31 & 71.95 & 62.50 & 69.00 \\
\midrule
Ours & -- & 63.64 & \textbf{\underline{53.40}} & \textbf{\underline{81.29}} & \textbf{\underline{81.82}} & \textbf{\underline{77.06}} & \textbf{\underline{74.39}} & \textbf{\underline{75.00}} & \textbf{\underline{71.00}} \\ 
\bottomrule
\end{tabular}
}
\end{table*}

\subsection{Evidence-Conditioned Audio Reasoning}
\label{sec:reason}

The core reasoning stage invokes Qwen-3-Omni-Instruct as the LALM backbone \(\mathcal{M}_{\text{A}}\), taking the original audio \(\mathbf{a}\), the question \(q\), the candidate options, and the structured evidence \(\mathcal{E}\) as joint input. Let \(r\) denote the current iteration index (\(r = 0\) for the initial attempt). At each iteration, the LALM generates a candidate:
\begin{align}
\hat{y}^{(r)} = \mathcal{M}_{\text{A}}\bigl(\pi_{\text{ans}},\mathbf{a},q,\{c_i\},\mathcal{E}, f^{(r-1)})
\label{eq:reason}
\end{align}
where \(f^{(r-1)}\) is the diagnostic feedback from the previous verification iteration (Section~\ref{sec:verify}). By injecting \(f^{(r-1)}\) into the prompt, the LALM is guided to attend to the specific weakness and produce a corrected response, rather than regenerating blindly.
The answer generation prompt \(\pi_{\text{ans}}\) instructs the LALM to:
(i)~decompose the question into intermediate sub-decisions;
(ii)~derive each intermediate conclusion by citing the corresponding evidence entries in \(\mathcal{E}\);
and (iii)~progressively compose these intermediate results into the final answer option, which is stated clearly in a predefined structured output format.By injecting pre-extracted evidence as explicit reasoning anchors, the LALM is steered to attend and attribute its decisions to question-relevant acoustic cues; moreover, our \textbf{prompted stepwise reasoning protocol}~\cite{wei2022chain} explicitly enforces intermediate sub-decisions grounded in evidence, strengthening the model's reasoning process beyond answer-only generation.

To reduce variance and improve reliability, we run the reasoning stage twice with different configurations (e.g., varied temperature and reordered evidence presentation), producing two candidate answers \(\hat{y}^{(1)}\) and \(\hat{y}^{(2)}\).
Both candidates, together with their reasoning traces, are passed to the verification stage.

\subsection{Verification And Output Arbitration}
\label{sec:verify}

The final stage ensures that the output is well-formed, logically consistent, and robustly grounded. Under the verification protocol \(\pi_{\text{ver}}\), \(\mathcal{M}_{\text{L}}\) performs three checks: \textbf{(a) Format compliance}—malformed outputs are repaired via rule-based post-processing to prevent parsing failures from being scored as incorrect; \textbf{(b) Reasoning--answer consistency}—the verifier checks for contradictions between the evidence cited in the CoT and the final answer, and any inconsistency triggers either answer correction or re-generation with targeted feedback highlighting the contradiction; and \textbf{(c) Dual-pass arbitration}—given two candidates \(\hat{y}^{(1)}\) and \(\hat{y}^{(2)}\), the verifier performs a comparative evaluation to select the final output:
\begin{align}
y^* = \mathcal{M}_{\text{L}}\bigl(\pi_{\text{ver}}, q, \mathcal{E}, \hat{y}^{(1)}, \hat{y}^{(2)})
\label{eq:verify}
\end{align}
When both candidates agree, the shared answer is adopted directly.When they disagree, the verifier re-examines the evidence chain and selects the candidate whose reasoning exhibits stronger evidence alignment and internal coherence.This mechanism provides an implicit self-consistency check~\cite{wang2023selfconsistency} without requiring expensive majority voting over many samples, and significantly reduces the rate of random errors in large-scale evaluation.

\section{Experiments}

\subsection{Setup}

We conduct experiments on the MMAR benchmark, reporting modality-wise performance under single-type and composite-type audio settings, where composite audios pose stronger interference and thus challenge question-relevant cue localization. We adopt end-to-end Qwen-3-Omni-Instruct as the primary baseline. We further perform ablations to verify that the gains come from distilling observations into question-tied evidence and enforcing a checkable reasoning procedure that can understand, integrate, and self-verify audio cues. All other settings (prompts, decoding configurations) are kept identical across runs.

\subsection{Evaluation Method}

All MMAR tasks are multiple-choice questions. We report \textbf{accuracy} (parsed option vs.\ ground truth) and the \textbf{rubric score} from the official MMAR evaluator, which measures faithfulness of the reasoning path to instance-level rubrics.

\subsection{Ablation Studies}
We conduct ablation studies on the 2026 Interspeech Audio Reasoning Challenge (Table~\ref{tab:ablation}) using Qwen-3-Omni-instruct as baseline to quantify the contribution of each component. Full pipeline achieves the best performance in both accuracy and rubrics, indicating that tool augmentation is beneficial when tool outputs are distilled into compact, decision-relevant evidence.

\textbf{w/o Evidence Integration.}
Removing integration leads to the largest degradation (5.6\% accuracy; 6.1 rubrics), \emph{even below the end-to-end baseline}. This highlights that tool use without evidence structuring can be harmful: directly injecting high-entropy observations introduces distracting context and increases option-mapping errors. Integration serves as a ``knowledge bridge'' that filters and aligns heterogeneous tool outputs into decision-ready evidence.

\textbf{w/o Observation.}
Disabling tool calls (1.8\% accuracy; 2.8 rubrics) collapses the pipeline back toward an end-to-end LALM that must ``listen and reason'' within a single hidden representation. Without \(\mathcal{O}\), the system loses access to explicit event/ASR/emotion/MIR cues (e.g., event timing, speaker content, affect, tempo), making it difficult to localize question-relevant segments and to resolve questions that hinge on fine-grained acoustic attributes or counting/temporal relations.

\textbf{w/o Verification.}
Removing verification decreases accuracy by 1.9\% and rubric scores by 1.5, showing that verification mainly improves robustness by correcting avoidable errors such as format violations and evidence--answer inconsistencies.

These results justify our design choice of prioritizing an auditable evidence chain over naive tool-use augmentation.

\subsection{Results}

Our method achieves \textbf{71.0\%} accuracy and a \textbf{63.0} rubric score on MMAR, ranking \textbf{5th} in the Agent Track. Compared with the end-to-end Qwen-3-Omni-Instruct baseline, we improve by \textbf{+2.3} accuracy points (68.7 $\rightarrow$ 71.0) and \textbf{+4.3} rubric points (58.7 $\rightarrow$ 63.0) under identical backbones and decoding settings. As shown in Table~1, EChO-Agent attains the \emph{best} average accuracy among all compared LALMs. The gains are most pronounced on composite (mixed-type) audio: converting raw tool observations into compact, question-tied evidence helps localize decision-critical cues rather than being distracted by salient but irrelevant segments. The rubric improvement further indicates a clearer, checkable reasoning process, where intermediate evidence is externalized as an auditable chain and verified for evidence--answer consistency. Ablations (Table~\ref{tab:ablation}) support this finding: removing evidence integration yields the largest drop, confirming its central role in reliable audio reasoning.

\begin{table}[htbp]
\centering
\caption{Ablation results of our proposed framework on the MMAR benchmark.Each ``w/o'' row removes a single component from the Full Pipeline.}
\label{tab:ablation}
\resizebox{\columnwidth}{!}{
\begin{tabular}{lcc}
\toprule
\textbf{Models} & \textbf{Accuracy} & \textbf{Rubrics} \\
\midrule
\textbf{Full Pipeline (Ours)} & \textbf{71.0} & \textbf{63.0} \\
\quad w/o Observation & 69.2 & 60.2 \\
\quad w/o Evidence Integration & 65.4 & 56.9 \\
\quad w/o Verification & 69.1 & 61.5 \\
\midrule
Qwen3-Omni-Instruct & 68.7 & 58.7 \\
\bottomrule
\end{tabular}
}
\end{table}

\section{Conclusion}

We proposed \textbf{EChO-Agent}, a tool-augmented framework that organizes audio reasoning into a \textbf{Tool $\rightarrow$ Evidence $\rightarrow$ Reason $\rightarrow$ Verify} pipeline with an auditable evidence chain. On MMAR, EChO-Agent achieves \textbf{71.0\%} accuracy and \textbf{63.0} rubric score, ranking \textbf{5th} in the Agent Track. Ablations confirm that structured evidence integration is the dominant contributor---raw tool outputs without filtering can degrade reasoning quality---while verification reduces avoidable last-mile errors. The granularity of sound-modality reasoning is currently bounded by perception tools: coarse event labels from YAMNet limit fine-grained sound understanding and remain a bottleneck on sound-centric questions. Future work will address this tool-induced granularity limit, together with tool uncertainty, cross-tool conflicts, and finer-grained temporal analysis.

\section{Generative AI Use Disclosure}
During the preparation of this manuscript, the authors used generative AI tools to polish the English language, improve readability, and assist with \LaTeX{} formatting. These tools were not used to generate any scientific claims, experimental results, or significant parts of the manuscript.

\bibliographystyle{IEEEtran}
\bibliography{mybib}

\end{document}